\documentclass[aps,showpacs]{revtex4}
\usepackage{epsfig}
\begin{document}

\title{Vector meson angular distributions \\in proton-proton collisions}
\author{Amand Faessler$^{\dagger )}$, C. Fuchs$^{\dagger )}$,  
M. I. Krivoruchenko$^{\dagger ,*)}$ and B. V. Martemyanov$^{\dagger ,*)}$}

\affiliation{$^{\dagger )}$Institut f\"{u}r Theoretische Physik$\mathrm{,}$ Universit\"{a}t 
T\"{u}bingen$\mathrm{,}$
Auf der Morgenstelle 14\\ D-72076 T\"{u}bingen, Germany\\
$^{*)}$Institute for Theoretical and Experimental Physics$\mathrm{,}$
B. Cheremushkinskaya 25\\ 117259 Moscow, Russia
}

\begin{abstract}
The resonance model is used to analyze the $\omega$- and $\phi$-meson 
angular distributions in proton-proton collisions at $\sqrt{s} = 2.83$
and $2.98$ GeV. The assumption of dominant contributions from  
$N^*(1720)\frac{3}{2}^+$ and $N^*(1900)\frac{3}{2}^+$
resonances which both have, according to the $\pi N$ scattering
multichannel partial-wave analysis and/or quark models predictions,  
dominant $p_{1/2}$ $N\omega$ decay modes yields the right pattern 
of the $\omega$ angular distribution at $\sqrt{s} = 2.83$ GeV. 
The angular distribution at $\sqrt{s} = 2.98$ GeV can be 
reproduced assuming the dominance of $N^*(2000)\frac{5}{2}^+$ and $N^*(1900)\frac{3}{2}^+$. The 
experimental $\phi$-meson angular distributions do not shown any
asymmetry which requires the existence of a 
massive negative-parity spin-half resonance. This resonance could be
identified with the $N^*(2090)\frac{1}{2}^-$. 

\end{abstract}
\pacs{13.60.Le, 14.20.Gk, 25.40-h}

\maketitle

The COSY-TOF collaboration \cite{COSYTOF} has recently measured the $\omega$-meson 
angular distribution in proton-proton collisions at an excess energy 
$\epsilon = 173$ MeV ($\sqrt{s} = 2m_p + m_{\omega} + \epsilon$) 
above the $\omega$ threshold. Earlier, angular distribution of
$\omega$- and $\phi$-mesons have been measured by 
the DISTO collaboration \cite{DISTO} at an energy excess of $\epsilon = 320$ MeV. 
The $\phi$-meson distribution is found to be consistent with a flat 
distribution, whereas the $\omega$-mesons at the both values of $\epsilon$ 
are peaked towards the beam directions.

The cross sections of vector mesons production enter as an input into
transport models for the dilepton emission in heavy-ion collisions 
\cite{FRA,GI,TUE}. 
The direct $\rho(\omega) \rightarrow e^{+}e^{-}$ decay channels give 
important contributions to the dilepton production. The detector efficiency depends 
thereby on the absolute value and direction of momentum of the dilepton pairs. Hence, the 
angular distribution of the vector 
mesons produced in nucleon-nucleon collisions affects the counting rate of 
the dilepton pairs which makes it necessary to keep the angular dependence of the 
production cross sections theoretically under the control. 

The dilepton spectra in heavy-ion collisions at $T = 1 - 5$ GeV/A were measured by 
the DLS collaboration 
\cite{DLS}. The HADES experiment \cite{HADES} is presently studying the dilepton 
spectra in the same energy region in great detail.

The pion production is the best known inelastic reaction in $pp$ collisions. A model 
independent partial-wave analysis of the experimental data has been  performed up to 
$\epsilon_{\pi} = 20$ MeV above the threshold \cite{FLA}. The pion angular distribution 
shows an increasing anisotropy with  energy.

The meson production in $pp$ collisions has previously been analyzed using the OBEP models \cite{OBEP}. 
The most of the realistic nucleon-nucleon interactions models have the $\Delta(1232)$ isobar included as
a special degree of freedom. It is naturally to assume that with increasing the energy other nucleon 
resonances also become important. The resonance model \cite{OSE,TEIS} as a first step treats matrix elements 
of the higher-mass resonance production as phenomenological constants fixed by fitting the
available inelastic cross sections. Such an approach is suitable to obtain estimates for the 
importance of various meson production channels and it allows to perform simulations of heavy-ion 
reactions in which the role of nucleon resonances is important. The dominance of nucleon resonances in 
the intermediate states is equivalent to taking nucleon-meson final-state interactions into account \cite{YAD}. 
The resonance model fits the $\rho,\;\omega,$ and $\phi$ total production cross sections rather well 
\cite{TUE,JOP,YAD,FAE}. 

The decay rates of nucleon resonances into the dilepton pairs and vector mesons with the same invariant mass 
coincide up to an overall kinematic factor. The $\Delta(1232)$ Dalitz decay is one of the major sources of dileptons in heavy-ion 
collisions at intermediate energies \cite{FRA,GI,TUE}. The first correct calculation of that decay was given 
only recently \cite{KRF}, whereas kinematically complete expressions for Dalitz decays of other high-spin 
resonances were given in \cite{AOP}. In this paper, we analyze the angular distributions of the $\omega$- and 
$\phi$-mesons in proton-proton collisions within the framework of the resonance model. 

\vspace{2mm}

According to the resonance model, the dominant contributions to the meson production cross sections 
originate from reactions $pp \rightarrow pN^*$ followed by subsequent decays $N^* \rightarrow p\omega 
(\phi)$. The first reaction takes place at $m_p + m_{N^*} \le \sqrt{s}$, while the second reaction with 
e.g. an $\omega$ production, takes place at $m_p + m_{\omega} \le m_{N^*}$. These inequalities can be 
combined to give $1.72 \le m_{N^*} \le 1.89$ GeV and $1.72 \le m_{N^*} \le 2.05$ GeV, respectively, for 
the COSY-TOF and DISTO experimental conditions. The $\phi$-meson production is sensitive to the mass
interval $1.96 \le m_{N^*} \le 2.04$ GeV.
 
Either a nucleon resonance falls into these intervals with its pole masses or, if not, it 
can have a Breit-Wigner leaking into the corresponding mass intervals. In the present work we will focus 
mainly on the first possibility. According to Particle Data Group \cite{PDG} the following resonances 
can be important: 
$N^*(1720)\frac{3}{2}^+$, $N^*(1900)\frac{3}{2}^+$, $N^*(1990)\frac{7}{2}^+$, and $N^*(2000)\frac{5}{2}^+$. 
The negative-parity resonances have masses below $1.72$ GeV, while $\Delta^*$ resonances do not contribute 
to the $\omega$ and $\phi$ production due to the isotopic symmetry.

At $\epsilon = 173$ MeV the resonances $N^*(1720)\frac{3}{2}^+$ and $N^*(1900)\frac{3}{2}^+$ appear at 
different kinematic conditions: In the $N^*(1720)$ decay, the $\omega$
is mainly produced at rest. In the $pp$ collisions, the 
$N^*(1720)$ is produced with a velocity $v \sim 0.26$. The $\omega$-meson angular distribution coincides 
therefore with the angular distribution of the $N^*(1720)$ resonance. 
The maximal angular momentum for the system $pN^*$ can be estimated as $\eta = p^*/m_{\pi}$ where 
$p^*$ is the c.m. momentum in the final state, $1/(2m_{\pi})$ is the strong interaction radius and 
$m_{\pi}$ is the pion mass. For $\epsilon = 173$ MeV, one has $\eta \sim 3.4$. One can expect that the 
partial-wave decomposition is truncated at a $pN^*$ orbital momentum $\sim 3$.

In the $N^*(1900)$ decay the $\omega$ is produced with velocity $v \sim 0.82$, whereas in the $pp$ 
collisions the $N^*(1900)$ is produced mainly at rest. The angular distribution of the 
$\omega$-mesons coincides therefore with the corresponding $\omega$ distribution in the $N^*(1900)$ 
resonance decay. This means that direction of the $\omega$ momentum is correlated with the resonance 
spin. The initial protons in the experiments \cite{COSYTOF,DISTO} are not polarized. The resonance spin 
is therefore correlated with the beam direction.

\vspace{2mm}

Let us first consider the $N^*(1900)$. The angular wave function of the $p\omega$ system, appearing due 
to a $J^P$ resonance decay, has the form
\begin{equation}
\Psi(\mathbf{n}_{\omega}) \sim \sum_{SL}A_{SL}C_{SS_{z}LL_{z}}^{JJ_{z}}C_{\frac{%
1}{2}\mu 1 \mu_{\omega} }^{SS_{z}}Y_{LL_{z}}(\mathbf{n}_{\omega}).
\label{pomega}
\end{equation}
The Clebsch-Gordan coefficients $C_{jj_{z} ll_{z}}^{JJ_{z}}$ are defined with the PDG phase 
conventions \cite{PDG}.
The partial-wave decomposition includes the total spins $S = 1/2$ and $3/2$ of the $p\omega$ system 
and the $p\omega$ orbital momenta $L$ such that $|J - S| \le L \le J + S$. Parity conservation 
gives a selection rule $P = (-1)^{L+1}$. The amplitudes $A_{SL}$ for different resonances are extracted 
from the $\pi N$ inelastic scattering data and/or predicted by the quark models (for a review see 
\cite{AOP}).

The $pN^*(1900)$ system appears close to the threshold in the $S$ state ($\eta \sim 0$). The initial 
protons have therefore the total angular momentum $j = 1$ or $2$ and positive parity. The 
state $1^+$ is forbidden by the Pauli principle, so the only possibility is $j^p = 2^+$ which is the
$^1{d}_2$ state for the initial protons. The higher states are classified in Table 1.

%%%%%%%%%
\begin{table}[tbp]
\caption{$j^{p}$ states of two protons at $j\ge 1$ and their $^{2s+1}l_{j}$
decompositions satisfying the Pauli principle.}
\label{labl2}
\begin{center}
\begin{tabular}{ll}
\hline\hline
$j^{p}$ & $^{2s+1}l_{j}$ \\ \hline
$j^{+}\;\mathrm{even}$ & $^{1}j_{j}$ \\ 
$j^{-}\;\mathrm{odd} $ & $^{3}j_{j}$ \\ 
$j^{+}\;\mathrm{odd} $ & $\mathrm{forbidden}$ \\ 
$j^{-}\;\mathrm{even}$ & $^{3}(j-1)_{j}\oplus {^{3}(j+1)_{j}}$ \\ 
\hline\hline
\end{tabular}
\end{center}
\end{table}
%%%%%%%%%%%

The transition amplitude $pp \rightarrow pN^*(1990)$ can be obtained constructing a rotational scalar 
out of the initial-state spherical harmonics $Y_{jj_{z}}(\mathbf{n}_{p})$, where $\mathbf{n}_{p}$ is the 
unit vector in direction of the proton beam. The final-state
$pN^*(1900)$ spin wave function reads:
\begin{equation}
\mathfrak{M}_{fi} \sim C^{jj_{z}}_{\frac{1}{2} \mu J J_{z}} 
                 Y_{jj_{z}}(\mathbf{n}_{p}). 
\end{equation}
Now, let the quantization axis of the angular momentum be parallel to the proton beam $\mathbf{n}_{p}$. The 
only non-vanishing component of the initial-state wave function is $j_z = 0$. The orbital momentum is 
perpendicular to the beam and its projection to the beam direction is equal to zero. The polarization 
matrix of a spin-$J$ resonance becomes then diagonal. The diagonal elements equal
\begin{equation}
\rho_{J_{z}J_{z}} = |C_{\frac{1}{2}-J_{z} JJ_{z} }^{j0}|^2, 
\label{polmat}
\end{equation}
and so $\rho_{J_z J_z} = 1/2$ at $J_z = \pm 1/2$ and, for a higher spin resonance, $\rho_{J_z J_z} = 0$ 
at $|J_z| \ge 3/2$. Eq.(\ref{polmat}) is valid for all even-$j$
positive-parity initial states. Spin-half resonances are therefore 
produced unpolarized near threshold, whereas higher spin 
resonances, including the $N^*(1900)$, are polarized.

The angular distribution of the $\omega$-mesons produced through the $N^*(1900)$ resonance decays can be 
found by weighting the polarization matrix (\ref{polmat}) with the modulus squared of the $p\omega$ angular 
wave function (\ref{pomega}) summed over the final-state $p\omega$ spin projections: 
\begin{equation}
\frac{d\sigma}{d\Omega}  \sim \sum_{J_{z}}\rho_{J_{z}J_{z}}\sum_{SS_{z}}
\left| \sum_{L}A_{SL}C_{SS_{z}LL_{z}}^{JJ_{z}}Y_{LL_{z}}(\mathbf{n}_{\omega})\right| ^{2}.
\label{nstar}
\end{equation}
The coefficient $C_{\frac{1}{2}\mu 1 \mu_{\omega} }^{SS_{z}}$ entering the $p\omega$ wave function
(cf. Eq.(\ref{pomega})) gives after summation over the $\mu$ and $\mu_{\omega}$ a decoherent sum over the
total spin and spin projection of the $p\omega$ system.

\begin{figure}[tb]
\par
\begin{center}
\leavevmode
\epsfxsize = 10cm \epsffile[25 60 573 465]{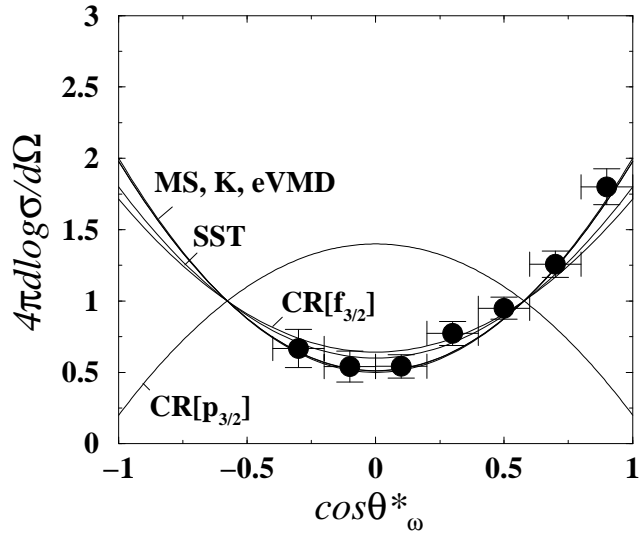}
\end{center}
\vspace{3mm}
\caption{The angular distribution of the $\omega$-mesons in the c.m. frame
of the colliding protons assuming the reaction goes through the $N^*(1900)\frac{3}{2}^+$ 
resonance. The experimental data from COSY-TOF \cite{COSYTOF} were 
obtained for an excess energy $\epsilon = 173$ MeV.  
The $p_{1/2}$ contribution $\sim 1 + 3cos^2{\theta^*_{\omega}}$ in the 
$p\omega$ decay of the $N^*(1900)$ is dominant according to the models MS \cite{MAN}, K 
\cite{KON}, and eVMD \cite{AOP}. The predictions of the models CR \cite{CR} and SST \cite{SST} 
are shown as well. 
}
\label{fig1}
\end{figure}

The partial-wave amplitudes $A_{LS}$ of the $N^*(1900) \rightarrow p\omega$ decay are extracted by 
Manley and Saleski \cite{MAN} from $\pi N$ multichannel partial-wave
analysis and/or quark model predictions by Koniuk \cite{KON}, Capstick
and Roberts \cite{CR}, and Stassart and Stancu \cite{SST} 
and/or predicted by the extended Vector Meson Dominance (eVMD) model \cite{AOP}. The resulting angular 
distributions are shown in Fig.\ref{fig1} together with the experimental data \cite{COSYTOF}. 

The models \cite{MAN,KON,AOP} predict a dominant $p_{1/2}$ wave with
an  angular dependence
$\sim 1 + 3cos^2{\theta^*_{\omega}}$ very close to the experiment. The models \cite{MAN,KON,AOP,SST}
are in very good agreement with the data. Capstick and Roberts
\cite{CR} report a vanishing 
$p_{1/2}$ amplitude and large errors for the $p_{3/2}$ and $f_{3/2}$ amplitudes. The dominance of the 
$f_{3/2}$ transition does not contradict to the model \cite{CR} and to the experimental data \cite{COSYTOF}. 
This amplitude is shown on Fig.\ref{fig1} as $CR[f_{3/2}$]. The $p_{3/2}$ transition can 
apparently be excluded.

If the $N^*(1900)$ resonance would be unpolarized, it could produce an isotropic $\omega$-meson 
distribution. The distributions plotted on Fig.\ref{fig1}, within the resonance model, are a consequence 
of the selection rules, the kinematic conditions, and the partial-wave content of the $N^*(1900)$ decay 
amplitude.

The $N^*(1900)$ width equals $\Gamma^{tot}_{N^*} = 500 \pm 80$
\cite{PDG}. In general, resonances are produced 
off-shell with invariant masses away from the pole mass. The off-shell $N^*(1900)$'s move 
with finite velocities which leads to a smearing of the $\omega$
angular distribution 
compared to the $N^*(1900)$ decays at rest. To estimate this effect, 
let us compare velocities of the $\omega$ in the $N^*(1900)$ rest frame and of the $N^*(1900)$ in the c. 
m. frame of two initial protons. For $\epsilon = 173$ MeV, we have $v_{\omega} = 0.46$ and $v_{N^*} = 0$ 
at $m_{N^*} = 1.9$ GeV and $v_{\omega} = 0$ and $v_{N^*} = 0.26$ at $m_{N^*} = 1.72$ GeV. The $\omega$ 
and $N^*$ velocities are equal $v_{\omega} = v_{N^*} = 0.23$ at $m_{N^*} = 1.76$ GeV, i.e. a 40 MeV 
above the $N^* \rightarrow p\omega$ decay threshold. The approximation in which the $N^*(1900)$ is 
treated as a particle at rest is therefore reasonable. Hence, 
we do not expect a large smearing effect due to 
the off-shell production of the $N^*(1900)$. 

%%%%%%%%%%%%%%%%%%%%%%%%%%%%%%%%%%%%%%%%%%%%%%%%%%%%%%%%%%%%%%%%%%%%%%%
%%%%%%%%%%%%%%%%%%%%%%%%%%%%%%%%%%%%%%%%%%%%%%%%%%%%%%%%%%%%%%%%%%%%%%%

%%%%%%%%%%%%%%%%%%%%%%%%%%%%%%%%%%%%%%%%%%%%%%%%%%%%%%%%%%%%%%%%%%%%%%%
%\subsection{$N^*(1720)$ at $\epsilon = 173$ MeV}
%%%%%%%%%%%%%%%%%%%%%%%%%%%%%%%%%%%%%%%%%%%%%%%%%%%%%%%%%%%%%%%%%%%%%%%%

\vspace{2mm}

Let us now discuss the $N^*(1720)$. The maximum $pN^*$ angular momentum at the excess energy $\epsilon = 
173$ MeV was estimated as $\sim \eta = 3.4$. Hence the $pN^*$ system can appear in orbitally excited 
states. The partial-wave content of the transition $pp \rightarrow
pN^{*}$ is unknown. A dominant  $S$-wave component can immediately be
excluded since it produces an isotropic $N^*(1720)$ and correspondingly
also an isotropic $\omega$ angular distribution. 
Let us check if $L=1$ and $L=2$ waves are allowed 
as dominant components.

\begin{figure}[tb]
\par
\begin{center}
\leavevmode
\epsfxsize = 10cm \epsffile[18 17 569 314]{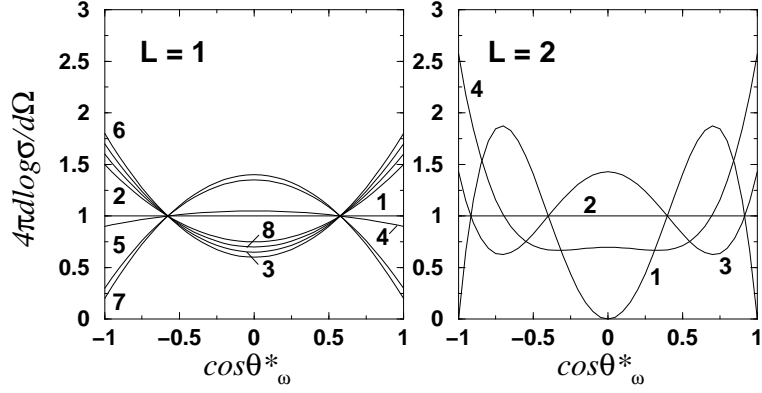}
\end{center}
%\vspace{3mm}
\caption{The normalized $L=1$ and $L = 2$ angular distributions of the $N^*(1720)\frac{3}{2}^+$ 
resonance in transitions ${^{2s + 1}l_{j}}{^{2S + 1}L_{j}}$ where $s$, $l$, and $j$ are the 
total spin, orbital momentum, and total angular momentum in the initial $pp$ 
state. $S$ and $L$ are the total spin and orbital momentum 
in the final $pN^*$ state. The numbers 1 to 8, 
attached to the $L = 1$ curves, correspond to the waves
${^{3}p_{0}}{^{3}P_{0}}$, ${^{3}p_{1}}{^{3}P_{1}}$, ${^{3}p_{2}{^{3}P_{1}}}$, 
${^{3}p_{1}{^{5}P_{1}}}$, ${^{3}p_{2}}{^{5}P_{2}}$, ${^{3}f_{2}}{^{3}P_{2}}$, 
${^{3}f_{2}}{^{5}P_{2}}$, and ${^{3}f_{3}}{^{5}P_{3}}$, respectively.
The numbers 1 to 4, 
attached to the $L = 2$ curves, correspond to the waves
${^{1}d_{2}}{^{3}D_{2}}$, ${^{1}s_{0}}{^{5}D_{0}}$, ${^{1}d_{2}}{^{5}D_{2}}$, and 
${^{1}g_{4}}{^{5}D_{4}}$.
}
\label{fig23}
\end{figure}

There exist eight $L=1$ amplitudes:
${^{3}p_{0}}{^{3}P_{0}}$, ${^{3}p_{1}}{^{3}P_{1}}$, ${^{3}p_{2}{^{3}P_{1}}}$, 
${^{3}p_{1}{^{5}P_{1}}}$, ${^{3}p_{2}}{^{5}P_{2}}$, ${^{3}f_{2}}{^{3}P_{2}}$, 
${^{3}f_{2}}{^{5}P_{2}}$, and ${^{3}f_{3}}{^{5}P_{3}}$ 
with quantum numbers 
${^{2s + 1}l_{j}}{^{2S + 1}L_{j}}$ where $s$, $l$, and $j$ are the total
spin, orbital momentum, and total angular momentum of the initial $pp$ 
state. $S$ and $L$ are the total spin and orbital momentum of the final $pN^*$ state. 
The angular distributions for the isolated transitions can be calculated from
\begin{equation}
\frac{d\sigma}{d\Omega} \sim \sum_{S_{z}L_{z}}
\left| C_{sj_{z}l0}^{jj_{z}}C_{SS_{z}LL_{z}}^{jj_{z}}Y_{LL_{z}}(\mathbf{n}_{\omega})\right| ^{2}.
\label{andist}
\end{equation}
They are plotted on Fig.\ref{fig23}. Since in the resonance rest frame the $\omega$-meson is produced at 
rest, the $\omega$ momentum has in the $pp$ c.m. frame the same direction as the $N^*(1720)$ 
momentum. Thus we have replaced in (\ref{andist}) $\mathbf{n}_{N^*}$ with $\mathbf{n}_{\omega}$. 

The wave Nos. 6, 8, 3, and 2 with quantum numbers
${^{3}f_{2}}{^{3}P_{2}}$, ${^{3}f_{3}}{^{5}P_{3}}$, ${^{3}p_{2}{^{3}P_{1}}}$, and 
${^{3}p_{1}}{^{3}P_{1}}$
resemble qualitatively the measured distribution, but yet 
no conclusions can be drawn on the actual importance of these waves. 
The ratio between maximal and minimal values of the differential cross section 
for those waves is below a factor of 3, whereas the experimental
ratio is close to a factor of 4. 
The resonance $N^*(1900)$ alone gives a stronger angular dependence,
being in slightly better agreement with the data.

There exist four $L=2$ transition amplitudes:
${^{1}d_{2}}{^{3}D_{2}}$, ${^{1}s_{0}}{^{5}D_{0}}$, ${^{1}d_{2}}{^{5}D_{2}}$, and 
${^{1}g_{4}}{^{5}D_{4}}$. The corresponding $\omega$-meson angular distributions are shown in 
Fig.\ref{fig23}. The ${^{1}g_{4}}{^{5}D_{4}}$ distribution resembles the experimental one, but has a 
plateau at $cos\theta ^*_{\omega} \sim 0$. The pattern of other waves is quite different from the 
observed one. 

The transitions $L=0$ and $L=1$  Nos. 1, 4, 5, and 7 and $L=2$ Nos. 1, 2, and 3 can apparently be 
excluded as dominant ones. The other four $P$-waves and the one $D$-wave transition can be large. 
These conclusions, however, are not stable against effects from the finite $N^*(1720)$ width of 
$\Gamma^{tot}_{N^*} = 150 \pm 50$ MeV \cite{PDG}:

The off-shell $N^*(1720)$ production results into an $\omega$ distribution representing a convolution of 
the $N^*(1720)$ and $\omega$ angular distributions. As we estimated, at $m_{N^*} = 1.76$ GeV the 
resonance and $\omega$ velocities are equal, so the $N^*(1720)$ tail $m_{N^*} \ge 1.76$ GeV generates 
roughly the $N^*(1720) \rightarrow N\omega$ decay distribution, whereas the invariant masses $1.72 \le 
m_{N^*} \le 1.76$ GeV correspond to one of the distributions plotted on Fig.\ref{fig23}. The domain of 
$40$ MeV above the $\omega$ threshold is not large compared to the resonance width. Hence the pattern of 
the decay $N^*(1720) \rightarrow N\omega$ is important. According to \cite{KON,AOP} the $p_{1/2}$ wave 
$\sim 1 + 3cos^2{\theta^*_{\omega}}$ of the $N^*(1720) \rightarrow N\omega$ decay is dominant. Other 
authors give negligible $\omega$ couplings for 
the $N^*(1720)$.

The $N^*(1720)$ production cross section is parameterized as an $S$-wave process \cite{TEIS}. This 
resonance provides $10 - 20\%$ of the total $pp\omega$ cross section \cite{JOP}. Such a value 
does not contradict to the $\omega$-meson angular distribution even in the sharp resonance limit.
The angular distribution data do not provide stringent constraints to the $S$-wave part of the 
$N^*(1720)$ production.

The $N^{*}(1900)$ is reliably predicted by quark models, but has currently a lower experimental 
status (**) than the $N^*(1720)$. The $N^*(1900)$ cross section is unknown and its 
contribution to the vector mesons production is set equal to zero \cite{JOP,YAD,FAE}. In the region 
$\sqrt{s} \sim 2.8$ GeV the vector meson production, according to
\cite{JOP,YAD,FAE}, is determined by 
tails of resonances of smaller masses and by the $N^{*}(1720)$. If we attribute the total cross 
section to the $N^*(1900)$ alone, we get $\sigma (pp \rightarrow pN^{*} (1900)) B(N^{*} (1900) 
\rightarrow p\omega) \sim 30 \; \mu b.$ Using the estimate of \cite{AOP} $\Gamma_{N^{*}p\omega} \sim 60$ 
MeV and the PDG value of $\Gamma_{N^{*}}^{tot} \sim 500$ MeV \cite{PDG}, one gets
$\sigma (pp \rightarrow pN^*(1900)) \sim 0.3$ mb at $\sqrt{s} \sim 2.8$ GeV. This is below the $s$-wave 
unitarity limit of $4\pi /k^2 \sim 4$ mb.

%%%%%%%%%%%%%%%%%%%%%%%%%%%%%%%%%%%%%%%%%%%%%%%%%%%%%%%%%%%%%%%%%%%%%%%
%%%%%%%%%%%%%%%%%%%%%%%%%%%%%%%%%%%%%%%%%%%%%%%%%%%%%%%%%%%%%%%%%%%%%%%

%%%%%%%%%%%%%%%%%%%%%%%%%%%%%%%%%%%%%%%%%%%%%%%%%%%%%%%%%%%%%%%%%%%%%%%
\begin{figure}[tb]
\par
\begin{center}
\leavevmode
\epsfxsize = 12cm \epsffile[18 17 569 314] {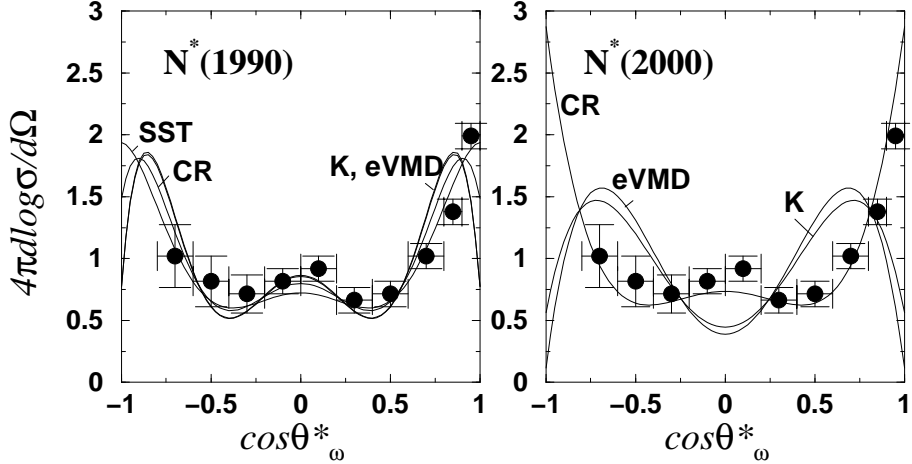}
\end{center}
%\vspace{3mm}
\caption{The angular distribution of the $\omega$-mesons in the c.m. frame
of the colliding protons assuming that the reaction proceeds through $N^*(1990)\frac{7}{2}^+$ 
and $N^*(2000)\frac{5}{2}^+$ decays. The models K 
\cite{KON}, CR \cite{CR}, SST \cite{SST}, and eVMD \cite{AOP} are compared
to the DISTO data at $\sqrt{s} = 2.98$ GeV \cite{DISTO}.
}
\label{fig34}
\end{figure}
%%%%%%%%%%%%%%%%%%%%%%%%%%%%%%%%%%%%%%%%%%%%%%%%%%%%%%%%%%%%%%%%%%%%%%%
\vspace{2mm}

Under the conditions of the DISTO experiment, 
we have $v_{\omega} = 0$ and $v_{N^*} = 0.35$ at $m_{N^*} = 1.72$ 
GeV and $v_{\omega} = 0.56$ and $v_{N^*} = 0$ at $m_{N^*} = 2.05$ GeV. The resonance and $\omega$ 
velocities at $\epsilon = 320$ MeV are equal to each other $v_{\omega} = v_{N^*} = 0.3$ at $m_{N^*} = 
1.8$ GeV.

The approximation $v_{\omega} \sim 0$ is good for the resonance $N^*(1720)$. The angular distribution 
due to the $N^*(1720)$ production is the same as in Fig.\ref{fig23}. The $L=2$ wave 
${^{1}g_{4}}{^{5}D_{4}}$ is the only wave that fits the DISTO $\omega$-meson data. 

The approximation $v_{N^{*}} \sim 0$ is good for the resonances $N^*(1900)$, $N^*(1990)$, and 
$N^*(2000)$. The angular distribution due to the $N^*(1900)$ production is the same as in 
Fig.\ref{fig1}. The pattern of the distribution is correct, but the $N^*(1900)$ does not develop a 
plateau at $cos \theta ^*_{\omega} \sim 0$ which is required by the DISTO data. The angular 
distributions of the $N^*(1990)$ and $N^*(2000)$ are shown on Fig.\ref{fig34}. The model \cite{CR} gives 
results close to the observations in both cases, the other models give reasonable angular 
dependences in the case of a $N^*(1990)$ dominance but stand in
contradiction to a possible $N^*(2000)$ dominance.

\vspace{2mm}

The $\phi$-meson angular distribution at $\sqrt{s} = 2.98$ GeV, i.e. 85 MeV above the $\phi$-meson 
production threshold, is found to be consistent with an isotropic distribution \cite{DISTO}. The 
maximal $\phi$ momentum can be estimated in the c.m. frame to be $p^{max}_{\phi} = 0.33$ GeV and the 
maximal $\phi$ orbital momentum relative to the $pp$ pair $\sim \eta_{\phi} = p^{max}_{\phi}/m_{\pi} = 
2.4$. The $S$-wave $\phi$-meson state allows $j^p = 1^{-}$ $pp\phi$ final states, whereas the higher 
orbital states are apparently excluded by the data which is somewhat surprising.

The $\omega - \phi$ mixing is sufficient to reproduces the $\phi$-meson production cross section 
at $\sqrt{s} = 2.98$ GeV \cite{FAE}. Such a mechanism, owing to the mass dependence of the $NN^* \omega$ 
coupling constants, yields in the nucleon resonance decays identical 
angular distributions for $\phi$- and $\omega$-mesons. 
To construct a $1^-$ final state out of the proton and a resonance in a 
$S$-wave, one needs negative-parity resonances. The well established negative-parity resonances 
$N^*(1535)\frac{1}{2}^-$, $N^*(1650)\frac{1}{2}^-$, and $N^*(1700)\frac{3}{2}^-$ with non-vanishing 
couplings to $\omega$-mesons have low masses and can only contribute
to the reaction through their Breit-Wigner tails. 
These resonances have dominant $s_{1/2}$ branchings \cite{KON,AOP} and
thus their tails would produce an isotropic cross section. The resonances $N^*(2090)\frac{1}{2}^-$ and 
$N^*(2080)\frac{3}{2}^-$ are listed by PDG \cite{PDG} with ratings (*)
and (**), respectively. Quark models predict several negative-parity 
states with masses around 2000 MeV but according to ref. \cite{CR} the
$s_{1/2}$ decays of these resonances are subdominant. 

The $J = \frac{1}{2}$ resonances in the $j^p = 1^-$ state give isotropic distributions. In the reference 
frame where the quantization axis of the angular momentum is parallel to the proton beam, the polarization 
matrix of the spin-$J$ resonance can be found to be
%%%%%%%%%%%%%%%%%%%%%%%%%%%%%%%%%%%%%%%%%%%%%%%%%%%%%%%
\begin{equation}
\rho_{J_z J_z} = \frac{2l + 1}{2j + 1} \sum_{s_z} |C^{js_z}_{ss_z l 0}C^{js_z}_{\frac{1}{2}s_z - J_z J 
J_z}|^2. 
\label{ueueue}
\end{equation}
%%%%%%%%%%%%%%%%%%
For $s=0$, we recover Eq.(3). Eq.(\ref{ueueue}) is valid for all $^{2s + 1}l_j$ states of two 
protons. The spin-half nucleon resonances are produced unpolarized. The diagonal elements of the 
$J = 3/2$ polarization matrix equal $\rho_{J_z J_z} =1/8$ and $3/8$ for $J_z = \pm 1/2$ and $\pm 3/2$, 
respectively. So, the $N^*(2090)\frac{1}{2}^-$ dominance in the $pp \rightarrow pp\phi$ or the dominance 
of any other spin-half nucleon resonance would result in a flat $\phi$-meson distribution. The 
high-spin resonances are polarized and their dominance can apparently be excluded (except for $J = 
\frac{3}{2}^-$ resonances with the dominant $s_{1/2}$ decay modes). 

Since the energy released in the process $pp \rightarrow pp\phi$ is small, it is instructive to compare 
the $\phi$ production with the near-threshold pion production which is well studied at $\epsilon_{\pi} 
\le 20$ MeV \cite{FLA}. The value of $\eta_{\pi} = p^{max}_{\pi}/m_{\pi}$ is small and varies in the 
limits $0.07 - 0.5$. In the limit $\eta_{\pi} \rightarrow 0$, only the $S$-wave survives, so the 
distribution must be symmetric. The large $S$-wave of the $NN\pi$ system e.g. at $\eta_{\pi} \sim 0.22$ 
implies an almost symmetric angular distribution which is reported \cite{FLA}. Within the resonance 
model, the dominance of the $S$-wave implies apparently the 
dominance near the threshold of tails of the negative-parity nucleon
resonances produced in an
 $S$-wave together with the nucleon. 
The importance of heavy resonances near the pion threshold is emphasized in 
\cite{RIS}. With increasing $\eta_{\pi}$, the pion distribution develops an anisotropy that becomes 
essential at $\eta_{\pi} \sim 0.5$. It can be attributed at least partially to the $\Delta(1232)$ which 
appears in the $S$-wave with another nucleon and generates the pion distribution $\sim 1 + 3cos^2 
\theta^*_{\pi}$, the same as the $\omega$-meson distribution in the $p_{1/2}$ wave with the proton, 
since the $\Delta(1232)$ has the same $p_{1/2}$ decay mode and the same spin as the $N^*(1900)$ 
resonance. Note that Eq.(\ref{nstar}) has no dependence on the constituent spins, only the total spin 
$S$ is important. At $\epsilon_{\pi} = 20$ MeV, the ratio between the $j^p = 1^-$ wave cross section 
that gives the flat distribution and the $j^p = 2^+$ wave cross section that gives the $\sim 1 + 3cos^2 
\theta^*_{\pi}$ distribution can be estimated from the angular distribution alone to give $\sim 3:2$ 
which is almost two times lower than the ratio of $\sim 5:2$ given by
\cite{FLA}. There is therefore 
space for contributions due to the asymmetry of other resonances and/or for higher orbital states of the 
$N\Delta(1232)$ system \cite{DMI}. In the $pp \rightarrow pp\phi$ reaction, while the $\phi$-meson is 
nonrelativistic, the maximal value of the orbital momentum $\sim \eta_{\phi} = p^{max}_{\phi}/m_{\pi}$ 
is several times larger than in the pion production studied in \cite{FLA}. These two near-threshold 
processes appear to be different with respect to the convergence rate
of the partial-wave expansion. 
The isotropic distribution observed by the DISTO collaboration can be interpreted as an evidence for 
a strong dominance of the $N^*(2090)\frac{1}{2}^-$ or any other spin-half negative-parity resonance. We 
would otherwise expect on the pure kinematic grounds an isotropic angular distribution starting from the 
$\omega$ production threshold of $\sqrt{s_{\phi}} = 2.90$ GeV up to 2.92 GeV where $\eta_{\phi} \le 1$ 
and an increased anisotropy beyond 2.92 GeV where higher orbital waves should come into play.
%%%%%%%%%%%%%%%%%%%%%%%%%%%%%%%%%%%%%%%%%%%%%%%%%%%%%%%%%%%%%%%%%%%%%%%
\begin{figure}[tb]
\par
\begin{center}
\vspace{3mm}
\leavevmode
\epsfxsize = 7cm \epsffile[57 46 514 479] {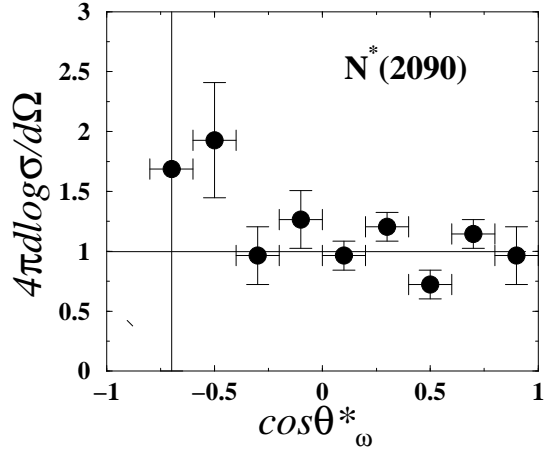}
\end{center}
\vspace{3mm}
\caption{The angular distribution of the $\phi$-mesons in the c.m. frame
of the colliding protons measured by the DISTO collaboration at $\sqrt{s} = 2.98$ GeV \cite{DISTO}
is compared to the flat distribution (solid line) generated by the dominant 
$N^*(2090)\frac{1}{2}^-$ contribution in the reaction mechanism.
}
\label{fig5}
\end{figure}
%%%%%%%%%%%%%%%%%%%%%%%%%%%%%%%%%%%%%%%%%%%%%%%%%%%%%%%%%%%%%%%%%%%%%%%
\vspace{2mm}

The angular differential cross sections provide sensitive tests for the reaction mechanisms. The present data
do not offer enough information for a complete partial-wave analysis. The resonance model which we discussed 
can be considered as first step to construct a microscopic $NN \rightarrow NN^*$ transition potential using
the $t$-channel exchange OBEP models or proposed recently an $s$-channel dibaryon exchange $NN$ interaction 
model \cite{kuku}.

The $N^*(1720)\frac{3}{2}^+$ and $N^*(1900)\frac{3}{2}^+$ resonances both have the dominant $p_{1/2}$ 
decay $N\omega$ modes. If velocities of these resonances in the
c. m. frame of two protons are not high 
compared to the $\omega$ velocities in the $N^*$ rest frames, which is a good approximation also for 
a significant off-shell spectral part of the $N^*(1720)$, the $N^* \rightarrow N\omega$ decays produce 
an angular distribution very close to the COSY-TOF data. The mechanism involving these two resonances 
could be the dominant one in the $\omega$ production at $\sqrt{s} = 2.83$ GeV. The several $P$ and the 
one $D$ wave of the $NN^*(1720)$ system generate also $\omega$ patterns close to the experimentally observed one.

At $\sqrt{s} = 2.98$ GeV a dominance of $N^*(1990)\frac{7}{2}^+$ and also 
$N^*(2000)\frac{5}{2}^+$ within the model \cite{CR} gives a reasonable description of the DISTO 
angular distribution data. The resonance $N^*(1900)\frac{3}{2}^+$ at rest generates still the right 
pattern but without a plateau at $cos\theta^*_{\omega} \sim 0$. The $N^*(1720)$ generates the right 
pattern through the ${^{1}g_{4}}{^{5}D_{4}}$ transition.

The $\phi$-meson data can be explained by the dominance of the $N^*(2090)\frac{1}{2}^-$. It is not 
clear, however, why higher orbital states of the $\phi$ and, respectively, high-spin nucleon resonances
which give anisotropic distributions should not be involved in the $\phi$-meson production.

We formulated also a general theorem according to which from every $^{2s + 1}l_j$ 
initial state of two unpolarized protons, spin-half nucleon resonances are produced at the threshold 
being unpolarized, whereas high-spin nucleon resonances are produced being polarized. The absence of any 
structure in the meson angular distribution indicates the possible dominance of a spin-half nucleon 
resonance, whereas the presence of a structure indicates the possible dominance of a $J \ge 3/2$ nucleon 
resonance.

A complete partial wave analysis requires angular distributions from
experiments with polarized beams. At the moment several competitive
mechanisms can explain the $\omega$-meson
production data while the $\phi$-meson distribution cannot be interpreted so 
easily. 

\vspace{2mm}

The authors are grateful to V. I. Kukulin for reading the manuscript and valuable remarks. M.I.K. and 
B.V.M. wish to acknowledge kind hospitality at the University of Tuebingen. This work is supported 
by DFG grant No. 436 RUS 113/721/0-1 and RFBR grant No.03-02-04004. 

%\newpage

\end{document}